\documentclass[10pt,conference]{IEEEtran}
\IEEEoverridecommandlockouts
\usepackage{cite}
\usepackage{amsmath,amssymb,amsfonts}
\usepackage{algorithmic}
\usepackage{graphicx}
\usepackage{textcomp}
\usepackage{xcolor}
\usepackage{tabularx}
\usepackage{subcaption}
\usepackage{url}
\usepackage{acro}[=v3]

\def\BibTeX{{\rm B\kern-.05em{\sc i\kern-.025em b}\kern-.08em
    T\kern-.1667em\lower.7ex\hbox{E}\kern-.125emX}}

\DeclareAcronym{CI}{
    short= CI,
    long= confidence interval
}
\DeclareAcronym{SD}{
    short= SD,
    long=Standard Deviation
}
\DeclareAcronym{ACCoRD}{
    short= ACCoRD,
    long=Acceleration of Compatibility and Commercialization for Open RAN
Deployments
}
\DeclareAcronym{POTOMAC}{
    short= POTOMAC,
    long=Platform for O-RAN Testing\, Orchestration\, and Management with AI Control
}
\DeclareAcronym{RSS}{
    short= RSS,
    long=Received Signal Strength
}
\DeclareAcronym{E2E}{
    short= E2E,
    long=End-to-End
}
\DeclareAcronym{SCS}{
    short= SCS,
    long=Subcarrier Spacing
}

\DeclareAcronym{RS}{
    short= R\&S,
    long=Rhode \& Schwarz
}
\DeclareAcronym{MCR}{
    short= MCR,
    long=Master Clock Rate
}
\DeclareAcronym{USRP}{
    short=USRP,
    long=Universal Software Radio Peripheral
}

\DeclareAcronym{SDR}{
    short=SDR,
    long=Software-Defined Radio
}

\DeclareAcronym{MIMO}{
    short=MIMO,
    long=Multiple-Input Multiple-Output
}

\DeclareAcronym{NI}{
    short=NI,
    long=National Instruments
}
\DeclareAcronym{SC}{
    short=SC,
    long=Software Community
}

\DeclareAcronym{ARFCN}{
    short=ARFCN,
    long=Absolute Radio Frequency Channel Number
}

\DeclareAcronym{RU}{
    short=RU,
    long=Radio Unit
}

\DeclareAcronym{DU}{
    short=DU,
    long=Distributed Unit
}

\DeclareAcronym{CU}{
    short=CU,
    long=Centralized Unit
}

\DeclareAcronym{O-CU}{
    short=O-CU,
    long=Open Centralized Unit
}

\DeclareAcronym{O-DU}{
    short=O-DU,
    long=Open Distributed Unit
}

\DeclareAcronym{O-RU}{
    short=O-RU,
    long=Open Radio Unit
}

\DeclareAcronym{DoD}{
    short=DoD,
    long=Department of Defense
}
\DeclareAcronym{MCS}{
    short=MCS,
    long=Modulation and Coding Scheme
}
\DeclareAcronym{EVM}{
    short=EVM,
    long=Error Vector Magnitude
}

\DeclareAcronym{CIC}{
    short=CIC,
    long=Cascaded Integrator-Comb
}

\DeclareAcronym{FR1}{
    short=FR1,
    long=Frequency Range 1
}

\DeclareAcronym{FR2}{
    short=FR2,
    long=Frequency Range 2
}

\DeclareAcronym{RF}{
    short=RF,
    long=Radio Frequency
}

\DeclareAcronym{SoC}{
    short=SoC,
    long=System-on-Chip
}

\DeclareAcronym{OTA}{
    short=OTA,
    long=Over-the-Air
}

\DeclareAcronym{UDP}{
    short=UDP,
    long=User Datagram Protocol
}

\DeclareAcronym{OAI}{
short=OAI,
long=OpenAirInterface
}
\DeclareAcronym{QAM}{
short=QAM,
long=Quadrature Amplitude Modulation
}
\DeclareAcronym{COTS}{
short=COTS,
long=Commercial Off-The-Shelf
}
\DeclareAcronym{LO}{
short=LO,
long=Local Oscillator
}
\DeclareAcronym{IF}{
short=IF,
long=Intermediate Frequency
}

\DeclareAcronym{PHR}{
short=PHR,
long=Power Headroom Report
}

\DeclareAcronym{FFT}{
short=FFT,
long=Fast Fourier Transform
}

\DeclareAcronym{TDD}{
short=TDD,
long=Time Division Duplex
}

\DeclareAcronym{Tx}{
short=Tx,
long=Transmit
}
\DeclareAcronym{Rx}{
short=Rx,
long=Receive
}
\DeclareAcronym{SIM}{
short=SIM,
long=Subscriber Identity Module
}
\DeclareAcronym{CQI}{
short=CQI,
long=Channel Quality Indicator
}

\DeclareAcronym{SNR}{
short=SNR,
long=Signal-to-Noise Ratio
}

\DeclareAcronym{RSRP}{
    short=RSRP,
    long=Reference Signal Received Power
}

\DeclareAcronym{SS-RSRP}{
    short=SS-RSRP,
    long=Synchronization Signal Reference Signal Received Power
}

\DeclareAcronym{NUC}{
    short=NUC,
    long=Next Unit of Computing
}
\DeclareAcronym{NR}{
    short=NR,
    long=New Radio
}
\DeclareAcronym{ML}{
    short=ML,
    long=Machine Learning
}
\DeclareAcronym{FPGA}{
    short=FPGA,
    long=Field Programmable Gate Array
}
\DeclareAcronym{SFDR}{
    short=SFDR,
    long=Spurious-Free Dynamic Range
}
\DeclareAcronym{AI}{
    short=AI,
    long=Artificial Intelligence
}
\DeclareAcronym{O-RAN}{
    short=O-RAN,
    long=Open Radio Access Network
}

\DeclareAcronym{UL}{
    short=UL,
    long=uplink
}

\DeclareAcronym{DL}{
    short=DL,
    long=downlink
}
\DeclareAcronym{DMRS}{
    short=DMRS,
    long=DeModulation Reference Signal
}
\DeclareAcronym{gNB}{
    short=gNB,
    long=gNodeB
}

\DeclareAcronym{UE}{
    short=UE,
    long=User Equipment
}

\DeclareAcronym{IoT}{
    short=IoT,
    long=Internet of Things
}

\DeclareAcronym{ADC}{
    short=ADC,
    long=Analog-to-Digital Converter
}

\DeclareAcronym{DAC}{
    short=DAC,
    long=Digital-to-Analog Converter
}

\DeclareAcronym{3GPP}{
    short=3GPP,
    long=3rd Generation Partnership Project
}

\DeclareAcronym{CCI}{
    short=CCI,
    long=Commonwealth Cyber Initiative
}

\DeclareAcronym{TS}{
    short=TS,
    long=Traffic Steering
}

\DeclareAcronym{RB}{
    short=RB,
    long=Resource Block
}

\DeclareAcronym{BW}{
    short=BW,
    long=Bandwidth
}

\DeclareAcronym{OFDM}{
    short=OFDM,
    long=Orthogonal Frequency Division Multiplex
}

\DeclareAcronym{PRB}{
    short=PRB,
    long=Physical Resource Block
}

\DeclareAcronym{TBS}{
    short=TBS,
    long=Transport Block Size
}

\DeclareAcronym{SLIV}{
    short=SLIV,
    long=Start and Length Indicator Values
}

\DeclareAcronym{srate}{
    short=srate,
    long=Sampling Rate
}

\DeclareAcronym{LOS}{
    short=LOS,
    long=Line of Sight
}

\DeclareAcronym{TFIG}{
    short=TFIG,
    long=Test and Integration Focus Group
}

\DeclareAcronym{BLER}{
    short=BLER,
    long=Block Error Rate
}

\DeclareAcronym{TCP}{
    short=TCP,
    long=Transmission Control Protocol
}

\DeclareAcronym{DL-TCP}{
    short=DL-TCP,
    long=Deep Learning-Based Transmission Control Protocol
}

\DeclareAcronym{DBM}{
    short=dBm,
    long=decibel-milliwatts
}

\DeclareAcronym{SINR}{
    short=SINR,
    long=Signal-to-Interference-plus-Noise Ratio
}

\DeclareAcronym{SS-SINR}{
    short=SINR,
    long=Synchronization Signal Signal-to-Interference-plus-Noise Ratio
}

\DeclareAcronym{QCI}{
    short=QCI,
    long=Quality of Service Class Identifier
}

\DeclareAcronym{RLC}{
short=RLC,
long=Radio Link Control
}

\DeclareAcronym{DUT}{
short=DUT,
long=Device Under Test
}

\DeclareAcronym{TIFG}{
short=TIFG,
long=Test and Integration Focus Group
}

\DeclareAcronym{BSR}{
    short=BSR,
    long=Buffer Status Report
}

\DeclareAcronym{ACK}{
    short=ACK,
    long=Acknowledgement
}

\DeclareAcronym{AM}{
    short=AM,
    long=Acknowledgement Mode
}

\DeclareAcronym{ORAN}{
  short = O-RAN,
  long = Open Radio Access Network
}

\DeclareAcronym{RAN}{
  short = RAN,
  long = Radio Access Network
}

\DeclareAcronym{Near-RT RIC}{
  short = Near-RT RIC,
  long = Near Real-Time RAN Intelligent Controller
}

\DeclareAcronym{Non-RT RIC}{
  short = Non-RT RIC,
  long = Non Real-Time RAN Intelligent Controller
}

\DeclareAcronym{RIC}{
  short = RIC,
  long = RAN Intelligent Controller
}

\DeclareAcronym{CU-CP}{
  short = CU-CP,
  long = Central Unit Control Plane
}

\DeclareAcronym{CU-UP}{
  short = CU-UP,
  long = Central Unit User Plane
}

\DeclareAcronym{E2SM}{
  short = E2SM,
  long = E2 Service Model
}

\DeclareAcronym{E2SM-KPM}{
  short = E2SM-KPM,
  long = E2 Service Model for Key Performance Measurement
}

\DeclareAcronym{KPM}{
  short = KPM,
  long = Key Performance Measurement
}

\DeclareAcronym{E2SM-RC}{
  short = E2SM-RC,
  long = E2 Service Model for RAN Control
}

\DeclareAcronym{RC}{
  short = RC,
  long = RAN Control
}

\DeclareAcronym{E2SM-NI}{
  short = E2SM-NI,
  long = E2 Service Model for Network Interface
}

\DeclareAcronym{E2SM-CCC}{
  short = E2SM-CCC,
  long = E2 Service Model for Cell Configuration and Control
}

\DeclareAcronym{5GC}{
  short = 5GC,
  long = 5G Core
}

\DeclareAcronym{5G}{
  short = 5G,
  long = Fifth Generation
}

\DeclareAcronym{SA}{
  short = SA,
  long = Standalone
}

\DeclareAcronym{NSA}{
  short = NSA,
  long = Non-Standalone
}

\DeclareAcronym{MAC}{
  short = MAC,
  long = Medium Access Control
}

\DeclareAcronym{PDCP}{
  short = PDCP,
  long = Packet Data Convergence Protocol
}

\DeclareAcronym{RRC}{
  short = RRC,
  long = Radio Resource Control
}

\DeclareAcronym{PHY}{
  short = PHY,
  long = Physical Layer
}

\DeclareAcronym{RSRQ}{
  short = RSRQ,
  long = Reference Signal Received Quality
}

\DeclareAcronym{CGI}{
  short = CGI,
  long = Cell Global Identifier
}

\DeclareAcronym{PCI}{
  short = PCI,
  long = Physical Cell Identifier
}

\DeclareAcronym{HO}{
  short = HO,
  long = Handover
}

\DeclareAcronym{TTT}{
  short = TTT,
  long = Time to Trigger
}

\DeclareAcronym{HOF}{
  short = HOF,
  long = Handover Failure
}

\DeclareAcronym{RLF}{
  short = RLF,
  long = Radio Link Failure
}

\DeclareAcronym{API}{
  short = API,
  long = Application Programming Interface
}

\DeclareAcronym{QoS}{
  short = QoS,
  long = Quality of Service
}

\DeclareAcronym{QoE}{
  short = QoE,
  long = Quality of Experience
}

\DeclareAcronym{KPI}{
  short = KPI,
  long = Key Performance Indicator
}

\DeclareAcronym{SLA}{
  short = SLA,
  long = Service Level Agreement
}

\DeclareAcronym{MLB}{
  short = MLB,
  long = Mobility Load Balancing
}

\DeclareAcronym{LB}{
  short = LB,
  long = Load Balancing
}

\DeclareAcronym{SON}{
  short = SON,
  long = Self-Organizing Network
}

\DeclareAcronym{CIO}{
  short = CIO,
  long = Cell Individual Offset
}

\DeclareAcronym{IP}{
  short = IP,
  long = Internet Protocol
}

\DeclareAcronym{DRL}{
  short = DRL,
  long = Deep Reinforcement Learning
}

\DeclareAcronym{ORAN Alliance}{
  short = O-RAN Alliance,
  long = Open Radio Access Network Alliance
}

\DeclareAcronym{ITU}{
  short = ITU,
  long = International Telecommunication Union
}

\DeclareAcronym{IEEE}{
  short = IEEE,
  long = Institute of Electrical and Electronics Engineers
}

\begin{document}

\title{Inter-DU Load Balancing in an Experimental Over-the-Air 5G Open Radio Access Network}
\author{
  \IEEEauthorblockN{
    \textsuperscript{$\dagger$}Fahim Bashar,
    \textsuperscript{$\dagger$}Asheesh Tripathi,
    \textsuperscript{$\dagger$}Mayukh Roy Chowdhury, 
    \textsuperscript{$\dagger$}Aloizio Da Silva,
    \textsuperscript{$\ast$}Alexandre Huff
  }
  \IEEEauthorblockA{\textsuperscript{$\dagger$}Commonwealth Cyber Initiative, Virginia Tech, USA, 
    \{fahimbashar, 
    asheesh, 
    mayukhrc, 
    aloiziops\}@vt.edu
  }
  \IEEEauthorblockA{\textsuperscript{$\ast$}Federal Technological University of Paraná, Brazil, 
    \{alexandrehuff\}@utfpr.edu.br
  }
}

\maketitle

\begin{abstract}
This paper presents the first ever fully open-source implementation of \ac{LB} in an experimental \ac{5G} \ac{NR} \ac{SA} network using \ac{ORAN} architecture. The deployment leverages the \ac{ORAN} \ac{SC} \ac{Near-RT RIC},  srsRAN stack, Open5GS core, and \acp{SDR}, with \ac{COTS} \acp{UE}. The implementation extends the srsRAN stack to support \ac{E2SM-RC} Style 3 Action 1 to facilitate \acp{HO} and adds \ac{MAC} \ac{DL} buffer volume reporting to srsRAN's \ac{E2SM-KPM}. The deployment demonstrates \ac{Near-RT RIC} closed-loop control where our \ac{MLB} xApp makes \ac{HO} decisions based on network load metrics for \ac{LB} between two \acp{O-DU} operating at different frequencies in the same band.
\end{abstract}


\section{Introduction}
\acl{LB} in cellular networks has evolved from a theoretical optimization problem to an operational necessity as \ac{5G} networks face unprecedented demands for guaranteed bit rates, ultra-reliable services, and energy savings. \ac{ORAN} promises to revolutionize this landscape through disaggregated architectures and standardized interfaces that enable \ac{TS} and \ac{MLB}. While simulation environments have advanced our understanding of \ac{MLB}\cite{ranfusion}, and \ac{AI}-driven approaches demonstrate \ac{MLB} performance improvements\cite{graph}, the majority of research is confined to simulation. Furthermore, commercial-grade \ac{TS} xApps like in \cite{rimedo} rely on proprietary tools and emulators that are not readily accessible to academic researchers. Although valuable, simulations and emulations cannot fully reflect the dynamic channel conditions and system-level challenges of real \ac{OTA} deployments. 

The disconnect between research advances and deployable solutions likely stems from the complexity of implementing \ac{ORAN} \ac{E2SM} and associated modifications in the \acs{RAN} stack. Open-source \ac{5G} stacks like srsRAN and OAI lack support for several of the service styles specified by the \ac{ORAN} Alliance in \ac{E2SM-RC} and \ac{E2SM-KPM}. This creates a significant bottleneck: researchers are constrained by open-source \ac{RAN} implementations that lack critical features for validating \ac{TS}/\ac{MLB} algorithms, while the industry gains limited value from academic work based on simulations and emulations.

This work presents the first ever \ac{OTA} experimental deployment of an \ac{ORAN}-based \ac{MLB} system, leveraging an open-source \ac{RAN} stack, core network, \ac{Near-RT RIC}, \acp{SDR}, custom xApp, and \ac{COTS} \acp{UE}. The deployment includes significant extensions to the srsRAN stack, adding support for \ac{E2SM-RC} Style 3 Action 1 for standardized \ac{HO} control via \acs{RC} messages. Additionally, \ac{E2SM-KPM} reporting was enhanced to incorporate \ac{MAC}-layer \ac{DL} buffer volume metrics, allowing for more informed and effective \ac{MLB} decisions.

\begin{figure}[!htb]
\centering
\includegraphics[width=\linewidth, height=8.0cm]{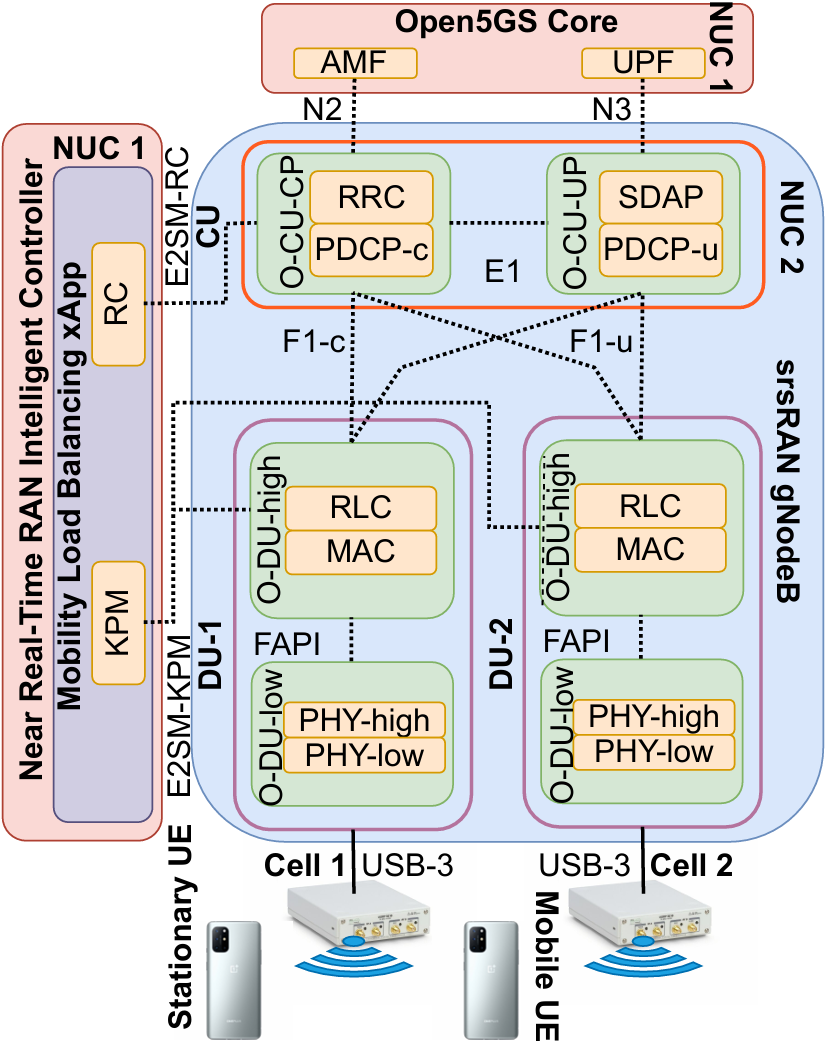}
\caption{End-to-End Architecture of the Experimental Setup}
\label{fig:system_diagram}
\end{figure}

\section{Implementation and Experimental Setup}

The experimental setup, as shown in Fig. \ref{fig:system_diagram} consists of three main components: a disaggregated \ac{RAN} implemented using modified srsRAN and Open5GS core, and \ac{ORAN} \ac{SC}'s \ac{Near-RT RIC} with our custom \ac{MLB} xApp developed at \ac{CCI} xG Testbed \cite{cci}. The \ac{RAN} includes an \ac{O-CU} connected to two \acp{O-DU} via F1 interfaces, supporting inter-\acs{DU} \ac{HO}.

The system uses Intel \acp{NUC} to host all software components. \ac{NUC} 1 runs Open5GS and the \ac{Near-RT RIC}, while \ac{NUC} 2 runs the modified srsRAN stack, including the \ac{O-CU}, both \acp{O-DU}, and interfaces to two \ac{USRP} B210s over USB 3.0. Each \ac{O-DU} connects to a dedicated \ac{SDR}, forming two overlapping \ac{5G} cells operating at different center frequencies within Band 78, each with 51 \acp{PRB} configured for a 20MHz bandwidth and 30kHz subcarrier spacing. This \ac{5G} deployment was benchmarked for \ac{OTA} performance as mentioned in \cite{bench}. The experiment uses two \ac{5G} \ac{COTS} \acp{UE}: a OnePlus 8T as the mobile \ac{UE} undergoing \ac{HO} and a OnePlus Nord as a stationary \ac{UE}. Both devices use programmable SysmoCom SIM cards, with \ac{DL} traffic generated using \texttt{iperf}. 

Our implementation extends srsRAN with two key enhancements to enable \ac{MLB} via xApp. First, we added support for xApp \ac{HO} control via \ac{E2SM-RC} Style 3 Action 1, which defines mobility control in connected mode\cite{oran_rc_spec}. The \acs{CU} was modified to process \ac{RC} messages and trigger internal mobility procedures based on the parameters received. We developed a mobility context interface to map \ac{RC} parameters and internal data structures used by the \ac{CU}. Second, we enhanced \ac{E2SM-KPM} reporting to support load-based mobility decisions. The \acs{DU} was modified to report \ac{MAC} \ac{DL} buffer volumes alongside existing \ac{PRB} utilization metrics, providing per-\ac{UE} and per-\ac{DU} measurements that are stored in the xApp as a 10-second rolling window. This approach offers a more holistic view of network load: \ac{PRB} utilization reflects current resource usage, while buffer volumes indicate data awaiting transmission.

The custom \ac{MLB} xApp is deployed in \ac{ORAN} \ac{SC}'s \ac{Near-RT RIC}, and a custom dashboard as shown in Fig. \ref{fig:dashboard} is used to visualize the \ac{DL} \ac{PRB} utilization, \ac{DL} throughput and \ac{MAC} buffer volume per \ac{DU} with low and high thresholds. In addition, it displays the individual \ac{UE} throughputs and the \ac{DU} on which they are currently attached.

\section{Demo}

\begin{figure}[!htb]
\centering
\includegraphics[width=\linewidth]{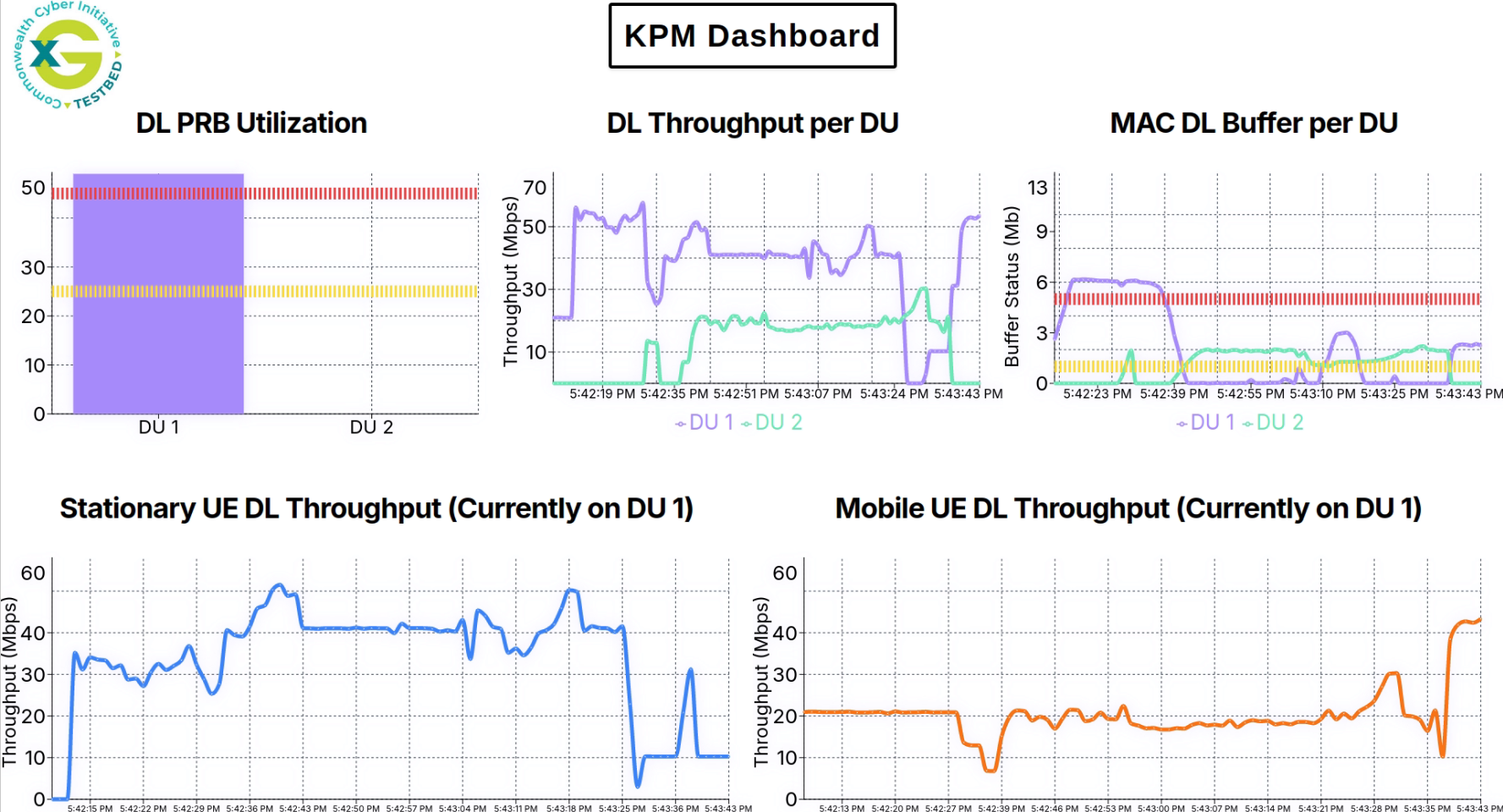}
\caption{Custom KPM Dashboard}
\label{fig:dashboard}
\end{figure}

The \ac{MLB} xApp implements an \ac{HO} triggering policy that requires \ac{DL} \ac{PRB} utilization and \ac{MAC} \ac{DL} buffer volume per \ac{DU}, to exceed or fall below the set thresholds. The xApp has subscribed to both \acp{O-DU} at a granularity of 1 second for both \ac{UE}-level and \ac{DU}-level metric tracking. The threshold values are determined through experimentation to best showcase the xApp's performance: for \ac{DL} \ac{PRB} utilization, the high threshold is set at 90\% and the low threshold at 50\% of the maximum available \acp{PRB}. The \ac{MAC} \ac{DL} buffer volume high and low thresholds are set at 5 Mbits and 1 Mbit, respectively. When these conditions are met, the xApp selects a target cell and issues a \ac{HO} command via the E2 interface. To avoid ping-pong \acp{HO}, a \ac{TTT} of 10 seconds is set. The xApp monitors each \ac{UE}’s association with an \ac{O-DU} and applies handover logic based on load conditions. If both \acp{UE} attached to \ac{DU}\textsubscript{1} and \ac{DU}\textsubscript{1}’s load remains above the high threshold for the \ac{TTT} interval, the xApp triggers an \ac{HO} of the mobile \ac{UE} to \ac{DU}\textsubscript{2}. If instead the \acp{UE} are on different \acp{O-DU} and \ac{DU}\textsubscript{1}’s load stays below the low threshold for the \ac{TTT} interval, the xApp issues an \ac{RC} to return the mobile \ac{UE} to \ac{DU}\textsubscript{1}. Fig.~\ref{fig:dashboard} illustrates a scenario with both \acp{UE} on \ac{DU}\textsubscript{1}: the mobile \ac{UE} had previously alternated between \ac{DU}\textsubscript{1} and \ac{DU}\textsubscript{2} as load changed, as reflected in the per \ac{DU} \ac{DL} throughput and \ac{MAC} \ac{DL} buffer volume plots. Initially only \ac{DU}\textsubscript{1} shows \ac{DL} throughput; after the handover, both \acp{DU} carry throughput, and the \ac{MAC} \ac{DL} buffer (initially above the high threshold, red dashed line) falls below the low threshold (yellow dashed line). When the mobile \ac{UE} returns to \ac{DU}\textsubscript{1}, throughput again appears only on \ac{DU}\textsubscript{1}. Throughout this scenario, the stationary \ac{UE}’s load is varied using \texttt{iperf} traffic.

We demonstrate an \ac{ORAN}-based, \ac{OTA}, closed-loop \ac{Near-RT RIC} platform enabling \ac{DU} \ac{LB} via our \ac{MLB} xApp. This validates the xApp's ability to make coordinated, network-wide optimization decisions based on a holistic view of \ac{RAN} conditions. The platform serves as a foundation for validating mobility-driven xApps for \ac{LB}, \ac{TS}, and energy savings, with potential for extension using machine learning techniques.

\section*{Acknowledgements}

This work was funded by \ac{CCI}. Visit \ac{CCI} at: \url{www.cyberinitiative.org}.

\bibliographystyle{IEEEtran}
\bibliography{references}

\end{document}